\def \be {\begin{equation}}

\def \half{\frac{1}{2}}
\def \ox {\otimes}

\def \real {{\bf R}}
\def \rf  {(\ref}
\def \eqn {equation}
\def \sln {solution}

\def \tfn {transformation}
\def \mi {matri}
\def\be{\begin{equation}}
\def\ee{\end{equation}}
\def\lbl{\label}

\documentstyle[12pt]{article}
\author {Ladislav   Hlavat\'{y} \small
\\ Faculty  of  Nuclear  Sciences  and Physical Engineering,
\\ Czech Technical University,
\\ B\v{r}ehov\'{a} 7, 115 19, Prague 1, Czech Republic. 
\\ E-mail: hlavaty@br.fjfi.cvut.cz} 
{\LARGE \bf\author{{\sc  Ladislav   Hlavat\'{y}}
\thanks{Postal   address:
B\v{r}ehov\'{a} 7, 115 19
Prague 1, Czech Republic. E-mail: hlavaty@br.fjfi.cvut.cz}
\\ {\it Department  of  Physics,}
\\ {\it  Faculty  of  Nuclear  Sciences  and
Physical Engineering}}

\title{
Principal models 
on a solvable group
with nonconstant metric}

\begin{document}
\maketitle
\abstract{Field equations for generalized principle models with nonconstant metric are derived and ansatz for their Lax pairs is given. Equations that define the Lax pairs 
are solved for the simplest solvable group. 
The \sln{} is dependent on one free variable that can serve as the spectral parameter. Painlev\'e analysis of the resulting model is performed and its particular \sln s are found}
\section{Introduction}
The Lax pair formulation for principal models with the trivial ($\delta_{ij}$) metric was published in \cite{zami:relinm} and Lax formulation of related $O(N)$-sigma models 
in \cite{pohlm:ihs}, \cite{lund:erci}. The models with the nontrivial but constant metric on the group manifold were investigated e.g. in \cite{cher:relinm}
 -- \cite{soch:igpcm}. The goal of this paper is to present an example of the principal model with a nonconstant metric on the group manifold having the Lax pair. The models with the nonconstant metrics we are aware of (e.g. \cite{mai:asa}, \cite{guka:ism}) are different from those investigated in this paper.

In the previous work  \cite{hla:halifx} we have given several conditions on the metric for $SU(2)$ models but we were not succesful in finding an example of the metric that is not constant on the group manifold and admits the Lax formulation of field equations.
Trying to find such an example, 
we have chosen much simpler group, namely the group of afine \tfn s of the real line -- denoted  here $Af(1)$. It is a nonabelian solvable two--dimensional Lie group that can be realized as the group of matrices
\be g  =  \left ( \begin{array}{cc} 
  a & b  \\
 0& 1
  \end{array} \right ), \ a>0,\ b\in \real.
\ee
It is diffeomorphic to $\real^2$ and one can choose the global coordinates $\theta_1, \theta_2$ on $Af(1)$ setting e.g. $a=\exp(\theta_1),\ b=\theta_2$.
The basis in the corresponding two--dimensional solvable Lie algebra 
can be chosen that the only nonvanishing structure coefficients are
\be {c_{12}}^2=-{c_{21}}^2=1 \lbl{feps}\ee
The coordinates of the left--invariant fields $J_\mu:=g^{-1}\partial_\mu g$ in this basis are $(\partial_\mu\theta_1,e^{-\theta_1}\partial_\mu\theta_2)$.
\section {Formulation of generalized principal chiral models}
Principal chiral models are 
given by the action
\be I[g]=\int d^2x \eta^{\mu\nu}L(J_\mu,J_\nu),\lbl{invaction}
\ee
where\be J_{\mu}:=(g^{-1}\partial_\mu g)\in{\cal L}(G), \lbl{amu}\ee
$g:\real^2\rightarrow G,\ \mu,\nu\in\{0,1\},\ \eta:=diag(1,-1)$ and $L$ is the Killing form on the corresponding Lie algebra ${\cal L}(G)$. 

An immediate generalization of the principal chiral models is obtained when one considers a general invertible bilinear form instead of the Killing \cite{cher:ie2das},\cite{soch:igpcm}.  
Next step of generalization \rf{invaction}) is introducing $G-$dependent symmetric bilinear forms. The generalized principal chiral models are then defined by the action
\be I[g]=\int d^2x L_{ab}(g)\eta^{\mu\nu}(g^{-1}\partial_\mu g)^a (g^{-1}\partial_\nu g)^b,
\lbl{action}\ee
where $L_{ab}(g)$ is invertible symmetric matrix $dim G\times dim G$ defined by the $G-$ dependent bilinear form $L(g)$ as 
\be L_{ab}(g):=L(g)(t_a\ox t_b), \ 
\ee
and $t_j$ are elements of a basis in the Lie algebra of the left-invariant fields. It is useful to consider the bilinear form $L(g)$ as a metric on the group manifold.
Lie products of elements of the  basis define the structure coefficients 
\be [t_a,t_b]={c_{ab}}^ct_c
\lbl{crfort}\ee
and in the same basis we define the coordinates of the field $J_\nu$
\be J_\nu=g^{-1}\partial_\nu g=J_\nu^bt_b 
\ee
that satisfy the Bianchi identities
\be \partial_\mu J_\nu-\partial_\nu J_\mu+[J_\mu,J_\nu]=0 \lbl{bianchi}\ee

Varying the action \rf{action}) w.r.t. $\eta:=g^{-1}\delta g$ we obtain \eqn s of motion for the generalized principal chiral models
\be \partial_\mu J^{\mu,a}+{\Gamma^a}_{bc}J_\mu^bJ^{\mu,c}=0,
\lbl{eqnsmot}\ee
where
\be {\Gamma^a}_{bc}:={S^a}_{bc}+{\gamma^a}_{bc},\lbl{Gamma}\ee
${S^a}_{bc}$ is the so called flat connection given by the structure coefficients
\be {S^a}_{bc}:=\half({C^a}_{bc}+{C^a}_{cb}),\ {C^a}_{bc}:=(L^{-1})^{ap} {c_{pb}}^qL_{qc}
\lbl{sdef}\ee
and ${\gamma^a}_{bc}$ are the Christoffel symbols for the metric $L_{ab}(g)$ on the group manifold
\be {\gamma^a}_{bc}:=\half (L^{-1})^{ad}(U_b L_{cd}+U_c L_{bd}- U_d L_{bc}). \lbl{christo}\ee
The vector fields $U_a$ in \rf{christo}) are defined in the local group coordinates $\theta_i$ as
\be U_a:=U_a^i(\theta)\frac{\partial}{\partial\theta_i} \ee
where the matrix $U$ is inverse to the matrix $V$ of vielbein coordinates
\be U_a^i(\theta):=(V^{-1})_a^i(\theta),\ \ V_i^a(\theta):=(g^{-1}\frac{\partial g}{\partial\theta_i} )^a.\ee
Note that the connection \rf{Gamma}) is symmetric in the lower indices
\be {\Gamma^a}_{bc}={\Gamma^a}_{cb}. \ee
For the group $Af(1)$
\be U_a=(\frac{\partial}{\partial \theta_1},e^{\theta_1}\frac{\partial}{\partial \theta_2}). \ee
\section{The Lax pairs}
In the paper \cite{soch:igpcm}, the ansatz for the Lax formulation of the generalized chiral models was taken in the form
\be [\partial_0+ P_{ab} J_0^b t_a+ Q_{ab} J_1^b t_a\ ,\ 
     \partial_1+ P_{ab} J_1^b t_a+ Q_{ab} J_0^b t_a]=0
\lbl{lpansatz}\ee
where $P,Q$ are two auxiliary $dim G\times dim G$ \mi ces. 
The ansatz \rf{lpansatz}) is a generalization of the Lax pair for Killing metric on the compact semisimple group $\L_{ab}=Tr(t_at_b)$ (in that case $Q$ and $P$ are multiples of the unit matrix).

{\em Necessary conditions that 
the operators in \rf{lpansatz}) form the Lax pair} for the \eqn s of motion \rf{eqnsmot}) are
\be \partial_0\,P=\partial_1\,Q,\ \ \partial_0\,Q=\partial_1\,P \lbl{conspq}\ee
\be (P_{bp}P_{cq}-Q_{bp}Q_{cq}) {c_{bc}}^a=P_{ab} {c_{pq}}^b ,
\lbl{condN}\ee
\be \half{c_{cd}}^a(P_{cp} Q_{dq}+P_{cq} Q_{dp})
= Q_{ab} {\Gamma^b}_{pq}.
\lbl{condNS}\ee
{\em If $Q$ is invertible}, that we shall assume in the following, then these conditions are {\em also sufficient.}
Note that the first two conditions are independent of the bilinear form $L$ so that one can start with solving the \eqn{} \rf{condN}) and then look for the bilinear forms $L$ that admit solution of the equation \rf{conspq},\ref{condNS}).
Moreover, as we have no {\em a priori} conditions for the first derivatives of the fields $g$ i.e. derivatives of $\theta_1,\theta_2$, we get from \rf{conspq})
\be \partial_{\theta_j}\,P=0,\ \partial_{\theta_j}\,Q=0 \lbl{constpq}\ee
\subsection{Solution of the equations \rf{conspq}--\ref{condNS}) for $Af(1)$}
The equations \rf{condN}--\ref{condNS}) where $a=1$ become linear equations for elements $P_{1b}$ and $Q_{1b}$ due to the fact that ${c_{pq}}^1=0$. It is rather easy to see that there is only trivial solution $Q_{11}=Q_{12}=0$ of the equations 
$Q_{1b} {\Gamma^b}_{pq}=0$ for constant nondegenerate metric $L$ (i.e. ${\Gamma^b}_{pq}={S^b}_{pq}$) so that  {\em there is no invertible matrix $Q$ that satisfy the equations  \rf{condNS}) for $Af(1)$ and the constant nondegenerate metric $L$}. 

On the other hand, there are solutions of the equations \rf{condN}-- \ref{condNS}) with $\det Q\neq 0$ for the nonconstant metric. In the following we shall present solutions for the diagonal metric
\be L(g)  =  \left ( \begin{array}{cc} 
  k_1(\theta_1,\theta_2) & 0  \\
   0 & k_2(\theta_1,\theta_2)
  \end{array} \right ).
\lbl{diagmetric}\ee

The equation \rf{condN}) for $Af(1)$ imply that matrix $P$ is of the form
\be P  =  \left ( \begin{array}{cc} 
  p_1 & 0  \\
  p_3 & p_2
  \end{array} \right )
\lbl{pmat}\ee
where 
\be p_2(p_1+1)=\det\ Q \lbl{psol}\ee

The linear equations for the elements $Q_{1b}$
given by \rf{condNS}) read
\[ Q_{1b}{\gamma^b}_{11}=0, \]
\be 2\, Q_{1b}{\gamma^b}_{12}+Q_{12}=0, \ee
\[ k_1 Q_{1b}{\gamma^b}_{22}-k_2 Q_{11}=0. \]
The condition $\det Q\neq 0$ requires that this system has a nontrivial solution. The nontrivial solvability of this system 
implies a system of quadratic equations for the Christoffel symbols that finally give a system of partial differential equations for the elements of the metric
\be (k_2  - k_{2, 1})k_{1, 1}- e^{2\theta_1} k_{1, 2}^2=0 \ee
\be (2 k_2 - k_{2,1})k_{1,2}+k_{1,1} k_{2,2}=0 \ee
\be (k_2  - k_{2, 1})(2 k_2 - k_{2,1}) +e^{2\theta_1} k_{1, 2} k_{2, 2} =0 \ee
where $k_{i,j}\equiv\frac{\partial k_i}{\partial\theta_j}. $
For every solution of this system we can get a solution of the equations \rf{condN}, \ref{condNS}) with $\det Q\neq0$. However, the equations \rf{conspq}) or \rf{constpq}) can be satisfied only if
\be k_1=K_1,\ k_2=K_2\exp(2\theta_1),\lbl{metres1}\ee
or 
\be k_1=K_1\exp(\theta_1),\ k_2=K_2\exp(\theta_1),\lbl{metres3}\ee
where $K_1, K_2$ are arbitrary nonzero constants.

For the case \rf{metres1}),
i.e. for
\be L(g) =  \left ( \begin{array}{cc} 
  K_1 & 0\\
  0 & K_2e^{2\theta_1}  \end{array} \right ),\ee
we get   
\be  P =  \left ( \begin{array}{cc} 
  1+\epsilon q_1 & 0\\
  c_1(1+\epsilon q_1)/q_1 & p_2  \end{array} \right ),\
  Q =  \left ( \begin{array}{cc} 
  q_1& 0\\c_1 &  \epsilon p_2 \end{array} \right ).
\lbl{res2}\ee
where $\epsilon=\pm 1,\ c_1,\ q_1,\ p_2$ are arbitrary constants $q_1\neq 0,\ p_2\neq 0$.
The equations of motion of this model read
\be \partial_\nu J^{1\nu}=0 \ee
\be \partial_\nu J^{2\nu}+J_\nu^1J^{2\nu}=0 \ee
that actually {\em represent the free model}
\be \partial_\nu\partial^\nu \theta_1  =0
\ee
\be \partial_\nu\partial^\nu \theta_2 =0 \ee
For the case \rf{metres3}) 
i.e. for 
\be L(g)  =  \left ( \begin{array}{cc} 
  K_1e^{\theta_1} & 0\\
  0 & K_2e^{\theta_1}  \end{array} \right ),\ee
the Lax pair is given by   
\be P =  \left ( \begin{array}{cc} 
\half & 0\\
\frac{q_2}{\kappa} & c_1\kappa  \end{array} \right ),\
  Q =  \left ( \begin{array}{cc} 
  0 & \frac{\kappa}{2}\\c_1 & q_2   \end{array} \right )
\lbl{res34}\ee
where $\kappa^2=-K_2/K_1\neq 0$ and $c_1,\ q_2$ are constants, $c_1\neq 0$. The equations of motion of this {\em nontrivial $Af(1)$ model with the nonconstant metrics} are
\be \partial_\nu J^{1\nu}+\half(J^1_\nu J^{1\nu}-\kappa^2 J^2_\nu J^{2\nu})=0 \lbl{af1eq1}\ee
\be \partial_\nu J^{2\nu}=0 \lbl{af1eq2}\ee
or
\be \partial_\nu\partial^\nu \theta_1 + \half(\partial_\nu\theta_1)(\partial^\nu\theta_1) -\half\kappa^2e^{-2\theta_1}(\partial_\nu\theta_2)(\partial^\nu\theta_2)=0
\lbl{af1model1}\ee
\be \partial_\nu\partial^\nu \theta_2 - (\partial_\nu\theta_1)(\partial^\nu\theta_2)=0 \lbl{af1model2}\ee
Inserting \rf{res34}) into \rf{lpansatz}) we get a linear combination of the equations \rf{af1eq1}), \rf{af1eq2}) containing the parameter $q_2$ and we find that without loss of generality we can set $q_2=0$. The Lax operators for the system \rf{af1eq1}), \rf{af1eq2}) then read
\be L_0=\partial_0+ \left ( \begin{array}{cc} 
  \half(J_0^1+\kappa J_1^2) & \lambda(J_1^1+\kappa J_0^2)\\0 & 0   \end{array} \right )
\lbl{l0}\ee
\be L_1=\partial_1+ \left ( \begin{array}{cc} 
  \half(J_1^1+\kappa J_0^2) & \lambda(J_0^1+\kappa J_1^2)\\0 & 0   \end{array} \right )
\lbl{l1}\ee
where $\lambda=c_1$ is free (spectral) parameter.

\section{Painlev\'e analysis and particular solutions}
To check the integrability of the model \rf{af1model1}),\rf{af1model2}) we can apply the usual Painlev\'e test \cite{ars:pconj}, \cite{wtc:ptest} i.e. to check the existence of the generic solution in the form of series with apropriate number of undetermined coefficients.

To be able to perform the test we must first convert the equations \rf{af1model1}),\rf{af1model2}) to the polynomial form. It can be done by the substitution $Z=\exp(\theta_1/2),\ W=\kappa\theta_2/2$ that transform the equations to the form
\be Z^3\partial_\nu\partial^\nu Z - (\partial_\nu W)(\partial^\nu W)=0 \lbl{af1model1zw}\ee
\be Z\,\partial_\nu\partial^\nu W - 2(\partial_\nu Z)(\partial^\nu W)=0 \lbl{af1model2zw}\ee
The singular point analysis is trivial as all possible leading powers for the above system are nonnegative. The analysis  around $Z=0$ can be done by setting $Z=1/Y$ and consequent singular point analysis of the system
\be Y\,\partial_\nu\partial^\nu Y - 2(\partial_\nu Y)(\partial^\nu Y)+Y^6(\partial_\nu W)(\partial^\nu W)=0 \lbl{af1model1yw}\ee
\be Y\,\partial_\nu\partial^\nu W + 2(\partial_\nu Y)(\partial^\nu W)=0 \lbl{af1model2yw}\ee
The singular solutions of this system have leading powers of $Y$ and $W$ (-1,3) and (-1,0). The resonances are $r=-3,-1,0,0$ and $r=-1,0,0,3$ respectively and compatibility conditions are satisfied. (In the former case they are trivial and in the latter case we have used a Mathematica code for the check.)

The polynomial form \rf{af1model1zw}), \rf{af1model2zw}) of the field equations is also convenient for obtaining particular solutions of the model. Indeed, choosing $Z(x_0,x_1)=z(y), W(x_0,x_1)=w(y)$ where $y=f(x_0,x_1)$ and $f$ 
solving the wave equation 
$\partial_\nu\partial^\nu f=0$ we obtain from \rf{af1model1zw}), \rf{af1model2zw}) a system of ODEs for $z,w$ that is rather 
easy to solve. We get
\[ z(y)=Ae^{By}+Ce^{-By},\]
\[ w(y)=D\pm\half(A^2e^{2By}+4ABCy-C^2e^{-2By}) \]
or 
\[ z(y)=y,\ w(y)=C \]
where $A,B,C,D$ are arbitrary real constants.

\section{Conclusion}
We have classified the generalized principal $Af(1)$ models 
that posses the Lax pair of the form \rf{lpansatz}). We have found that 
no such model with constant metric $L_{ab}$ in the action \rf{action}) exists and that there are just two models with the nonconstant diagonal metric. One of them is the two--component free model and the other one is given by the equations of motion \rf{af1model1}), \rf{af1model2}). A particular solution of the latter model was obtained and the Painlev\'e analysis confirms its integrability.

\end{document}